\RequirePackage{lineno}
\documentclass[aps,prr,twocolumn,letterpaper,superscriptaddress,showpacs]{revtex4-1}
\usepackage[colorlinks,linkcolor=blue,anchorcolor=blue, citecolor=blue,urlcolor=blue,]{hyperref}
\usepackage{graphicx}
\usepackage{CJK}
\usepackage{verbatim}
\usepackage{physics}
\usepackage{mathptmx}
\usepackage{color}
\begin{document}

\begin{CJK*}{UTF8}{bsmi}

\title{Mottness induced superfluid phase fluctuation with increased density}

\author{Zi-Jian Lang (\CJKfamily{gbsn}郎子健)}
\affiliation{Tsung-Dao Lee Institute \& School of Physics and Astronomy, Shanghai Jiao Tong University, Shanghai 200240, China}

\author{Fan Yang (\CJKfamily{gbsn}杨帆)}
\affiliation{School of Physics, Beijing Institute of Technology, Beijing 100081, China}

\author{Wei Ku (\CJKfamily{bsmi}顧威)}
\altaffiliation{corresponding email: weiku@sjtu.edu.cn}
\affiliation{Tsung-Dao Lee Institute \& School of Physics and Astronomy, Shanghai Jiao Tong University, Shanghai 200240, China}
\affiliation{Key Laboratory of Artificial Structures and Quantum Control (Ministry of Education), Shanghai 200240, China}

\date{\today}

\begin{abstract}
Recent observation of diminishing superfluid phase stiffness upon increasing carrier density in cuprate high-temperature superconductors is unexpected from the quantum density-phase conjugation of superfluidity.
Here, through analytic estimation and verified via variational Monte Carlo calculation of an emergent Bose liquid, we point out that Mottness of the underlying carriers can cause a stronger phase fluctuation of the superfluid with increasing carrier density.
This effect turns the expected density-increased phase stiffness into a dome shape, in good agreement with the recent observation.
Specifically, the effective mass divergence due to ``jamming'' of the low-energy bosons reproduces the observed nonlinear relation between phase stiffness and transition temperature.
Our results suggest a new paradigm, in which unconventional superconductivity in some strongly correlated materials is described by physics of bosonic superfluidity, as opposed to pairing-strength limited Cooper pairing.
\end{abstract}
\maketitle
\end{CJK*}

\section{Introduction}
\begin{figure}
	\begin{center}
		\includegraphics[width=\columnwidth]{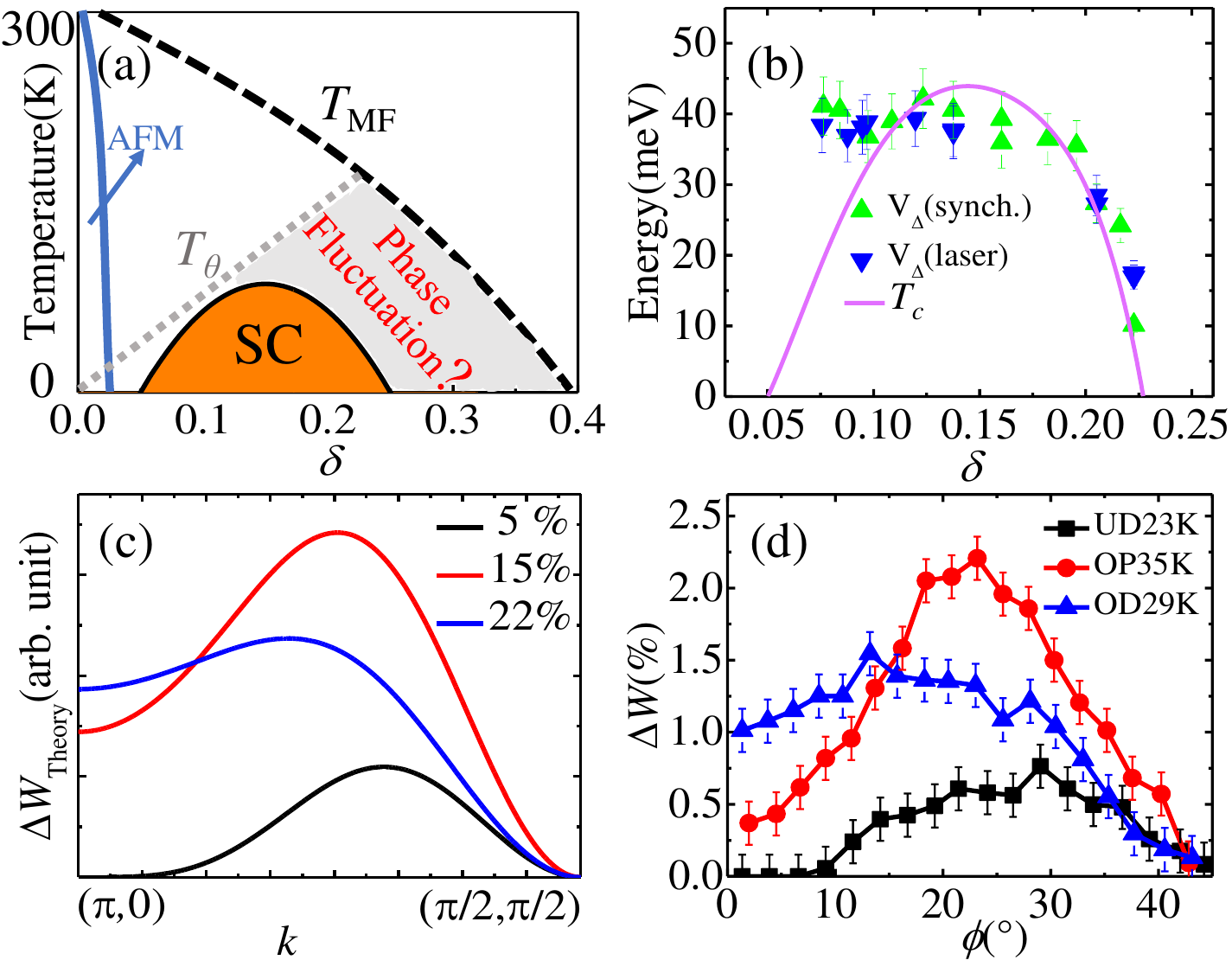}
	\caption{Indications of the essential role of phase fluctuation even in the overdoped cuprates. (a) Schematic phase diagram of hole-doped HTSC.  Beyond the antiferromagnetic (AFM) phase at very low doping, the superconductivity exists in a ``dome'' region of the phase diagram, much smaller than the one predicted by the VMC calculation.  This suggests that below the classical phase coherent temperature, $T_\theta$, additional phase fluctuation must exist in the region in gray. (b) Observed superconducting gap size (triangles) near momentum ($\pi/2$,$\pi/2$)~\cite{Vishik,Shen} displays a doping dependence very different from that of superconducting temperature (purple line).  Normalized spectral weight transfer $\Delta W$ from condensed EBL (c)  and ARPES~\cite{Kaminski} (d)both show a $d$-wave form distinct from the $\cos(k_x)-\cos(k_y)$ form of the superconducting order parameter, against expectation of BCS-like theories.}
	\label{fig1}
	\end{center}
\end{figure}

Superfluid phase stiffness is known to suffer from low density, due to the canonical conjugation between carrier density and phase $\phi$ $\Delta n\Delta\phi \sim \hbar$\cite{Emery1}.
Since the density fluctuation is very limited at low density, the phase must then fluctuate significantly, leaving only a small portion in the well-defined phase of the superfluid component.
This quantum mechanical limitation on superfluid density provides a natural explanation~\cite{Doniach,Emery1} for the observed reduction of the superconducting transition temperature $T_c$ in the underdoped cuprates [cf. $\delta<15\%$ of Fig.~\ref{fig1}(a)].
As shown schematically in Fig.~\ref{fig1}(a), in the underdoped region ($\delta<15\%$), $T_c$ decreases as the phase stiffness is weakened at lower carrier density (gray dotted line).
In that case the superconducting transition temperature $T_c$ would be controlled by the phase stiffness, as opposed to the strength of the pairing.

Unexpectedly, recent measurement~\cite{Bozovic1} of penetration depth $\lambda$ in the overdoped cuprates ($\delta>15\%$) found a surprising reduction of superfluid stiffness upon increasing carrier density.
Correspondingly, the low-temperature superfluid phase stiffness ($\propto\lambda^{-2}$) forms a dome shape against doping, qualitatively different from the simple proportionality to carrier density~\cite{Zaanen} expected in the pairing strength-limited BCS theory.
Furthermore, the low-temperature phase stiffness even scales with $T_c$ in the overdoped regime, obeying the same universal Uemura relation~\cite{Uemura1,Uemura2,Uemura3} well-known in the underdoped regime.
Since this relation implies a phase coherence-limited superfluidity, this new data apparently reveals that even in the overdoped side, the dominant physics is still the phase fluctuation.
Nonetheless, it is not obvious at all why superfluid stiffness should reduce at high carrier density.

In fact, several previous studies already raised similar doubts against the common belief of BCS-like amplitude fluctuation in the overdoped cuprates, and faced the same puzzle how the superfluidity can suffer at higher density.
As shown in Fig.~\ref{fig1}(b), ARPES observed\cite{Vishik, Shen} a nearly doping independent superconducting gap $\Delta$ near momentum $(\pi/2,\pi/2)$ over a wide doping range($\sim$7-20\%), in great contrast to the expected proportionality to $T_c$ in the BCS theory.
Similarly, Fig.~\ref{fig1}(d) shows that even earlier, the observed momentum-dependence of superconducting gap-induced weight transfer deviates qualitatively from the expected $d_{x^2-y^2}$ form of the superconducting order parameter, also from the underdoped regime \textit{all the way to the overdoped regime}.
Both observations suggest more commonality between the underdoped regime and the overdoped one than previously expected.
Consistently, variational Monte-Carlo (VMC) calculations~\cite{Paramekanti1,Paramekanti2,Taoli}, which ignore phase fluctuation, all found that the pairing still exists beyond 25\% doping [black dashed line in Fig.~\ref{fig1}(a)].
In this case, the entire superconducting dome would be deep inside the pairing region of the phase diagram, and thus the demise of superconductivity in the overdoped region must be controlled by additional phase fluctuation of unknown origin.

So, the key scientific question is how it is possible to host such a strong superfluid phase fluctuation at such high carrier density?
Particularly in the overdoped cuprates, how can the phase quantum-fluctuates even stronger with increased carrier density, as to wipe out superconductivity at roughly $\delta > 25\% $?
Why is this critical doping level seemingly universal across different families of cuprates and the new nickelate superconductors~\cite{Danfeng1,Danfeng2}?
What determines this special doping level?
What is the origin of the striking similarity between the underdoped and overdoped regime, especially near the quantum critical points (the end of the dome) $\delta\sim 5\%$ and $25\%$?
Finally, what essential nature do these puzzles reveal about the unconventional high-temperature superconductivity in the cuprates?

Here we show that all these important questions can be answered naturally in the scenario of the emergent Bose liquid~\cite{WeiKu1,WeiKu2,Jiang,Hegg2021,Yue2021,Zeng2021}.
Essentially, the high-energy local Coulomb repulsion that prevents double occupation of carriers in each lattice site at low energy (so-called ``Mottness'') leads to a rather large hard core of the superfluid carriers at even lower temperature.
Such a large hard core would unavoidably cause a serious jamming in carriers' kinetic motion, and in turn damage the superfluid phase coherence at higher density, a trend opposite to the common lore.
Below we demonstrate this general mechanism of stronger superfluid fluctuation at higher density via simple close-pack estimation and Gutzwiller approximation through statistical counting and variational Monte Carlo, all showing a consistent destruction of superfluidity at around 25-30\% carrier density.
Specifically for the cuprates and nickelates, this classically intuitive mechanism explains the robust upper limit of doping level in the entire families of cuprate and nickelate superconductors.
Combined with the persistent consistency of optical~\cite{Zeng2021}, transport~\cite{Hegg2021,Zeng2021}, quasi-particle~\cite{WeiKu1,Jiang,Yue2021}, and superconducting~\cite{WeiKu2} properties of the exact same model with corresponding experimental observations, our results reinforce the notion of a new paradigm that the high-temperature superconductivity in the cuprates (and likely nickelates) is dominated by physics of bosonic superfluidity, as opposed to pairing-strength limited Cooper pairing.

The contents of this paper are organized as follows. Section~\ref{model} introduces the emergent Bose liquid model and the basic assumptions. In Section~\ref{results}, we discuss the main results of this study. Section~\ref{EBL_check} gives the connection of our model to the case of the cuprate superconductors.

\section{Model}
\label{model}
\begin{figure}
	\setlength{\belowcaptionskip}{-0.4cm}
	\begin{center}
	\includegraphics[width=\columnwidth]{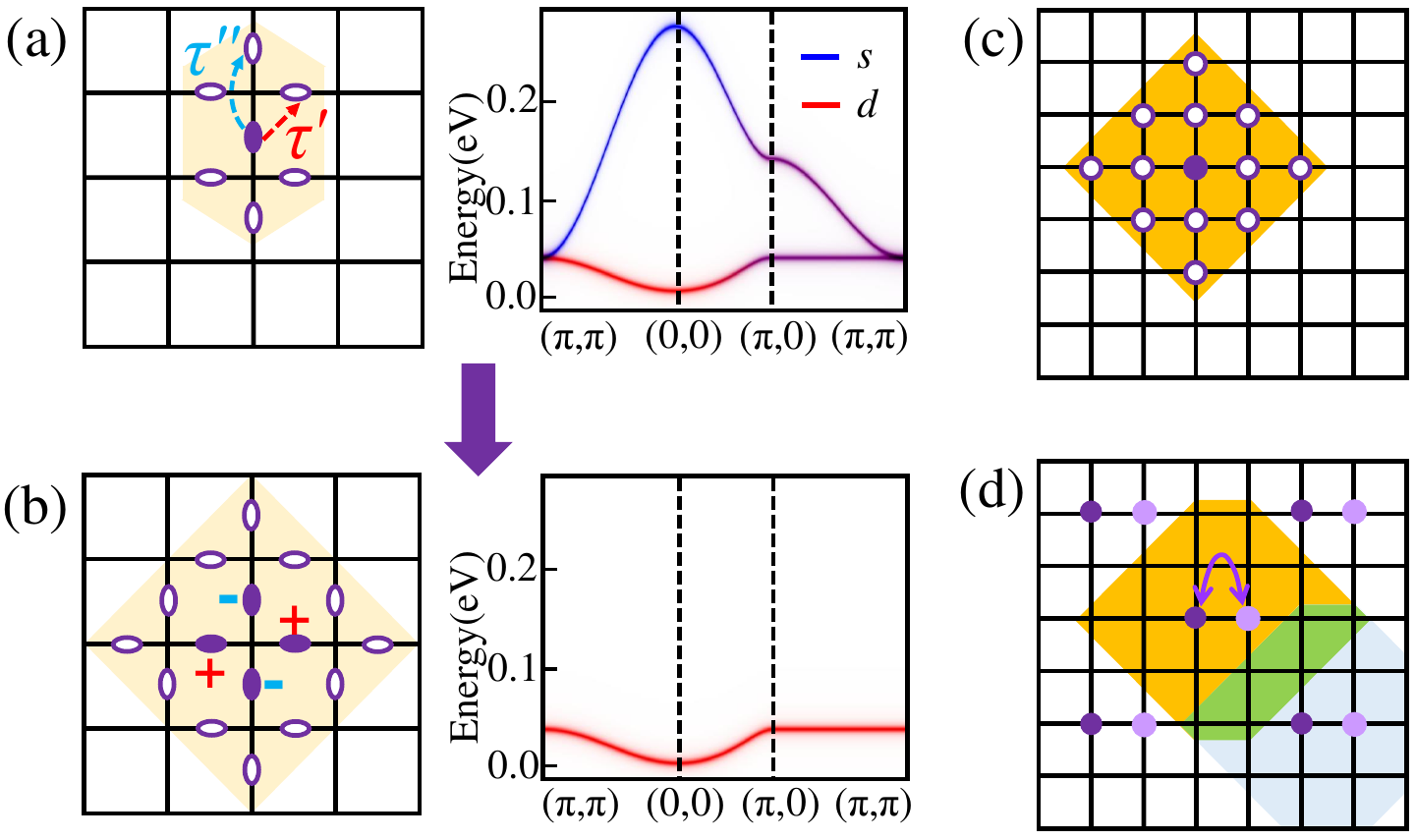}
	\end{center}
	\caption{Model of an emergent Bose liquid. (a) Hopping of a bond-centered boson (solid ellipse) to nearest $\tau^\prime$ and next nearest $\tau^{\prime\prime}$ neighboring bonds (empty ellipse), giving rise to a two-band dispersion (right panel), colored in red/blue to reflect the $d$- and $s$-wave nature.  (b) $d$-wave boson corresponding to the low-energy effective one-orbital description with one band dispersion (right panel), requiring 16 forbidden bonds.  (c) Extended hard-core constraint among site-centered $d$-wave bosons, up to third neighboring sites.  (d) Simple close-pack estimation of the highest boson density with minimal non-zero kinetic energy allowing hopping within two-sites on average.}
	\label{fig2}
\end{figure}
In the emergent Bose liquid (EBL) model~\cite{WeiKu1,WeiKu2,Jiang,Hegg2021,Yue2021,Zeng2021}, the charge carriers are assumed to be mostly in a tightly bound two-body state involving the nearest neighboring atoms, below some energy scale, say 150meV, due to various high-energy physics, for example, bi-polaronic correlation~\cite{Alexandrov}, two-dimensional short-range anti-ferromagnetic correlation~\cite{Dean,Sawatzky1} and/or cancellation of topological spin current~\cite{Weng}.
(For example, in cuprates this bosonic two-body state would consist of doped holes of opposite spin residing in neighboring Cu sites, using the half-filled Mott insulator with strong short-range antiferromagnetic correlation as reference.)
Correspondingly, the probability for the fermionic carriers in an unbound one-body state is relatively low below this energy scale.
Therefore, most physical properties of the system are dominated by the emerged bond-centered bosonic carriers [c.f. solid ellipse in Fig.~\ref{fig2}(a)] that reside in the corresponding bond lattice of the original fermionic lattice [black cross in Fig.~\ref{fig2}(a)].
Note that this assumption of EBL does not require a complete depletion of the probability for low-energy fermion (or fermionic spectral weight), but only its overall smallness.
(More detailed considerations on low-energy fermions are available in Appendix~\ref{App1}.)

In addition, in many strongly correlated materials, intra-atomic electronic repulsion is very strong, well above the energy scale of this effective low-energy model, resulting in an exclusion of double occupation~\cite{Hirsch,Emery2,doubleoccupacy} (so-called ``Mottness'') in the low-energy sector of the fermionic carriers.
This gives rise to an `extended hard-core constraint' of EBL, namely forbidden occupation of the vicinity of a boson by another, as illustrated by the shadowed area in Fig.~\ref{fig2}(a).

The kinetic process of the EBL is dominated by the pivoting motion~\cite{WeiKu1} in the emerged bond-centered lattice.
For the specific case of a square lattice of fermions, for example the Zhang-Rice singlets~\cite{Zhang1988} in the cuprates, the bond lattice is a checkerboard lattice in Fig.~\ref{fig2}(a) with \textit{two} orbitals in the unit cell corresponding to the vertical and horizontal bonds:
\begin{equation}
\label{model1}
H = \sum_{<l,l^\prime>} \tau_{ll^\prime} b_{l}^\dagger b_{l^\prime}\quad \text{(+ constraint)}
\end{equation}
where $b_{l}$ denotes the annihilation of a boson located at \textit{bond} $l$, and $\tau_{ll^\prime}=\tau^\prime$ or $\tau^{\prime\prime}$ is the fully dressed kinetic process involving the nearest and second-nearest neighboring bonds.
In this study, we employ the exact same set of parameters $\tau^\prime$ and $\tau^{\prime\prime}$ previously extracted from ARPES dispersion data~\cite{Yoshida2006,Yoshida2009,WeiKu1} (cf. Appendix~\ref{parameters}.)

It is important to emphasize that the low-energy effective Hamiltonian of EBL is under the above-mentioned extended hard-core constraint, which prevents phase separation related to clustering of the emergent bosons.
At low temperature, it also helps to establish the superfluid phase stiffness in the superfluid phase.
As demonstrated below, this constraint is the key factor that generates the surprising physical behavior of interest in this study.

Now, since the focus of our study is on the phase stiffness of superfluidity, which involves only the lowest energy excitation, it is advantageous to  further integrate out irrelevant (higher-energy) kinetic degree of freedom of Eq.~\ref{model1}.
Figure~\ref{fig2}(a) shows that within the bosonic band structure, the lower-energy band is dominated by $d$-wave symmetry due to the positive $\tau^\prime$.
We proceed to decouple the $d$-wave boson from the $s$-wave one via a canonical transformation.
Practically, this is performed approximately via construction of the $d$-wave bosonic Wannier orbital corresponding to the lower-energy band in Fig.~\ref{fig2}(a), and then representing the effective Hamiltonian in this decoupled subspace.
(See Appendix~\ref{App2} for more details.)
Essentially, this is equivalent to assuming the formation of a well-defined lowest energy quasi-particle with a $d$-wave structure responsible for the superfluidity.

Finally, we arrive at the resulting lowest-energy $one$-orbital Hamiltonian that controls directly the physics of phase fluctuation,
\begin{eqnarray}
\label{Honeband}
\tilde{H} = \sum_{<i,i^\prime>} \tilde{\tau}_{ii^\prime} d_{i}^\dagger d_{i^\prime}\quad \text{(+ interactions \& constraint)}
\end{eqnarray}
corresponds to a \textit{larger} $d$-wave Wannier orbital, denoted by $d^\dagger_i$ creation operator, now centered at \textit{sites} $i$ in the \textit{original square lattice} [cf.: Fig.~\ref{fig2}(b)].
Note that due to the shift of the Wannier center, the $d$-wave boson is not strictly local, but instead with a power-law decay similar to the Zhang-Rice singlet~\cite{Zhang1988}.
Correspondingly, the $d$-wave boson can now hop beyond the first and second neighbors with further renormalized $\tilde{\tau}_{ii^\prime}$ given in Tab~\ref{tabA2}.

Most importantly, these low-energy superfluid related $d$-wave bosons now acquire an even more extended hard-core constraint [shadowed area in Fig.~\ref{fig2}(b)] over 16 bonds.
As highlighted in Fig.~\ref{fig2}(c), this constraint forbids occupation of 12 surrounding \textit{sites} by other low-energy $d$-wave bosons.

\section{Mottness induced superfluid phase fluctuation with increased density}
\label{results}

Concerning the main puzzles of the strong phase fluctuation and diminishing superfluidity density at high carrier density, we will now show that the EBL also provides a simple resolution through its ``extended hard-core constraint''.
Such a large extended hard-core implies that the essential kinetic processes will be easily suppressed at moderate density due to blocking and jamming between the low-energy $d$-wave bosons.
Figure~\ref{fig2}(d) shows a simple estimation of the special case, in which each of the bosons reach minimum non-zero mobility on average, being able to hop back and forth between two sites.
The corresponding density, 2 holes (1 boson) / 8 atomic sites $= 25\%$, marks the approximate maximum doping level of superfluidity, above which the $d$-wave bosons become nearly impossible to move and thus unable to maintain phase coherence.
Interestingly, this 25\% coincides very well with the experimentally observed end of the superconducting dome in overdoped cuprates in general~\cite{Dagotto}.
Therefore, this mechanism offers a simple and natural explanation of the key puzzle of strong phase fluctuation in the overdoped cuprates.

\begin{figure}
	\setlength{\abovecaptionskip}{-0.3cm}
	\setlength{\belowcaptionskip}{-0.5cm}
	\begin{center}
	\includegraphics[width=10cm]{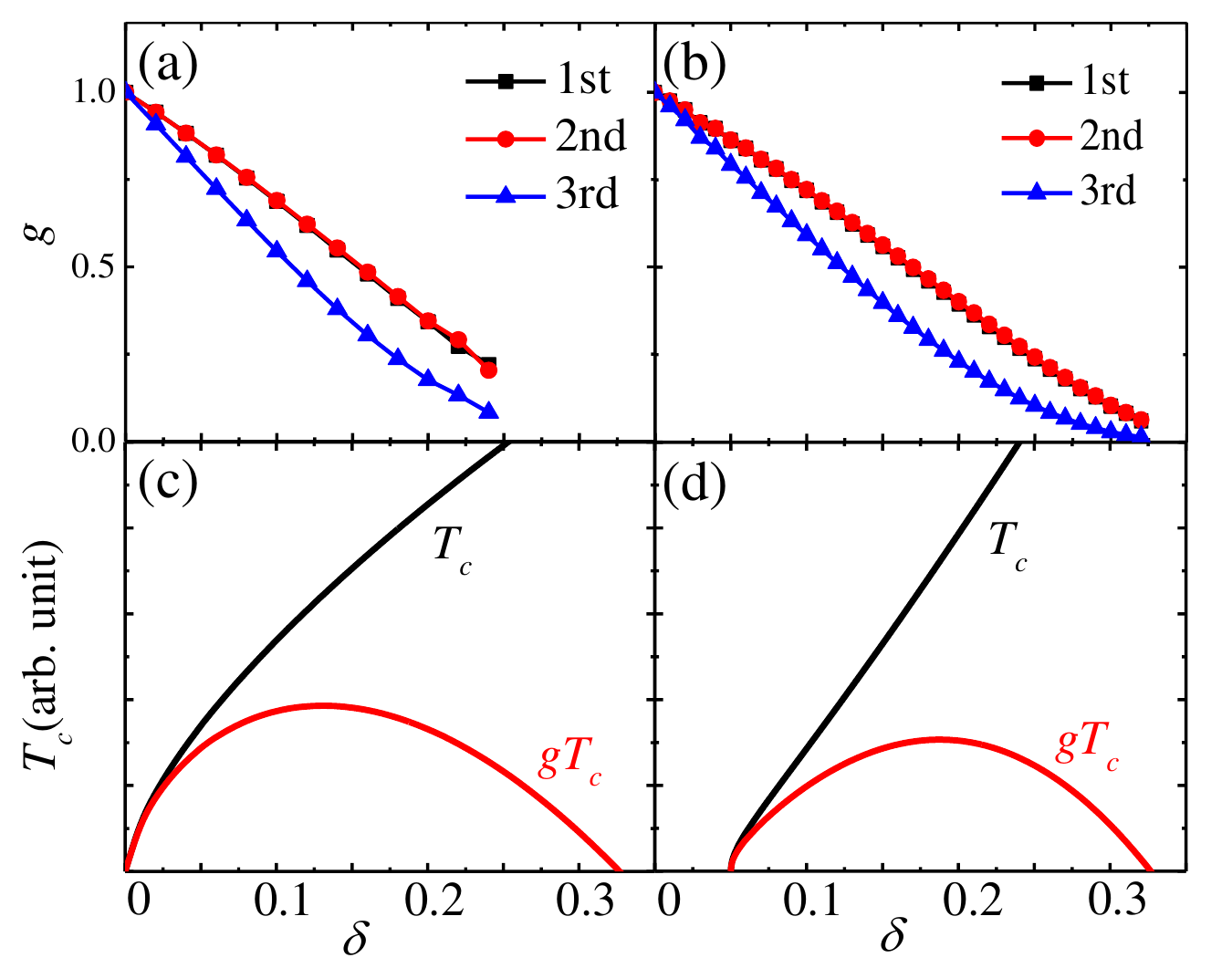}
	\end{center}
	\caption{Gutzwiller $g$-factor of the hopping parameters, calculated from (a) statistical counting and (b) VMC.  (c) Theoretical (black line) and renormalized (red line) transition temperature $T_c$ estimated from standard free Bose gas and the Gutzwiller approximation.  (d) same as (c) but with realistic mass from Ref.~\cite{WeiKu1,WeiKu2} containing additional mass divergence around $\delta\sim 5\%$ doping.}
	\label{fig3}
\end{figure}
We employ the well-known Gutzwiller $g$-factor approximation~\cite{Gutzwiller,Kotliar2} $\tilde{\tau}_{ii^\prime}\rightarrow g_{ii^\prime}\tilde{\tau}_{ii^\prime}$ as the simplest way to capture this renormalization of the kinetic process.
The $g$-factors are calculated via two numerical approaches (See Appendix~\ref{gfac}).
First, we count statistically the probability of each hopping under the constraint at various doping level.
Figure~\ref{fig3}(a) shows that the resulting $g$-factors are approximately $g \sim 1 - \delta/0.3$.
Second, we calculate the $g$-factor using VMC based on a noninteracting BEC wavefunction under the constraint.
The resulting $g$-factors in Fig~\ref{fig3}(b) resemble Fig~\ref{fig3}(a) very well.
As expected, both results decrease rapidly and become rather small beyond 25\% doping and diminish around 30\%.

We now demonstrate the ``dome'' shape of $T_c$ with our simple picture.
Estimated from the standard BEC~\cite{BEC} of a uniform boson gas, the condensation temperature $T_c\propto n^{2/3}/m^*$ is a simple function of density $n$ and effective mass $m^*$.
Obviously, following $\tilde{\tau}$, $1/m^*$ and $T_c$ are also renormalized by $g$.
Figure~\ref{fig3}(c) plots $T_c\propto \delta^{2/3}$ (in black) and the renormalized $T_c \rightarrow g T_c$ (in red).
Indeed $g T_c$ is strongly suppressed at higher doping and eventually vanishes around 30\%, where the average $m^*$ diverges.
This $\delta\sim30\%$ upper bound of superfluidity is in excellent agreement with Tl$_2$Ba$_2$CuO$_{6+\delta}$~\cite{Bangura} and well consistent with the common $\sim25\%$ limit of typical cuprates superconductors.
If one further incorporates the previously proposed doping dependent bosonic mass that diverges at around 5\% due to a level crossing~\cite{WeiKu2}, the $g T_c$ in Fig.~\ref{fig3}(d) reproduces the experimentally observed dome shape quite well.
In short, the phase fluctuation can indeed grow in the overdoped regime when the suppression of kinetic processes overcomes the increasing density.

\begin{figure}
	\begin{center}
		\includegraphics[width=\columnwidth]{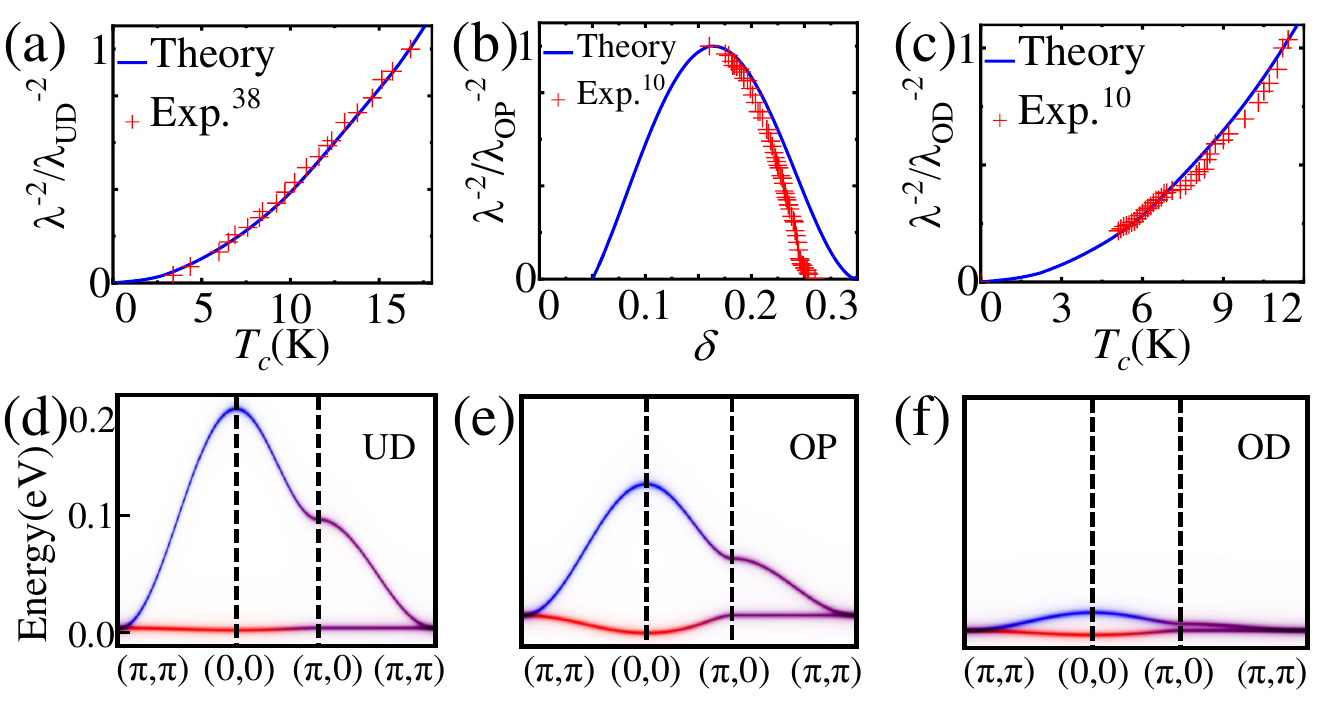}
	\end{center}
	\caption{(a)Correlation between low-temperature superfluid stiffness $\lambda^{-2}$ and superconducting transition temperature $T_c$ near 5\% quantum phase transition point~\cite{Broun,WeiKu2}, normalized by $\lambda_{UD}^{-2}$ at $T_c=17$K.  (b)Superfluid phase stiffness $\lambda^{-2}$ vs doping level $\delta$ showing a dome shape~\cite{Bozovic1}, normalized by $\lambda^{-2}_{OP}$ at $\delta=0.16$.  Theoretical overestimation in the overdoped region is expected from the Gutzwiller treatment.  (c) same as (a) for the overdoped region~\cite{Bozovic1}, normalized by $\lambda^{-2}_{OD}$ at $T_c=11$K.  (d)-(f) Renormalized band structure of EBL near $\sim 5\%$, $\sim 15\%$, and $\sim 25\%$ doping, showing similar effective mass divergence of low-energy band near transition points, but with distinct higher energy features.
    }
	\label{fig4}
\end{figure}

Similarly, the low-temperature limit of the phase stiffness $\lambda^{-2}$ would also suffer from the jamming reduced kinetic process.
Phase stiffness $\lambda^{-2}\propto n_s / m^*$ is proportional to the superfluid density $n_s$.
Near the above-mentioned mass divergence dictated quantum phase transition points, $n_s\propto (1/m^*)^\beta \delta $ must vanish with $1/m^*$ with a exponent $\beta>0$, leading to $\lambda^{-2} \propto  (1/m^*)^{1+\beta} \delta$.
Figure~\ref{fig4}(b) shows that this phase stiffness indeed forms a dome shape, vanishing at around $\delta\sim 5\%$ and $\sim 30\%$.
Particularly, due to the extended hard-core constraint of the EBL, the phase stiffness reduces in the overdoped regime against the growing density, as observed in recent experiments~\cite{Bozovic1,Zaanen}.
(Our results' deviation from experiment merely reflects the above-mentioned overestimation of phase coherence at high doping in our simple Gutzwiller treatment.)

Furthermore, the effective mass divergence of EBL has an important physical consequence in the relation between $\lambda^{-2}$ and $T_c$~\cite{WeiKu2}.
Estimated from $T_c\propto \delta^{2/3}/m^*$ for a free boson gas, one finds $\lambda^{-2} \propto T_c^{1+\beta}$.
Therefore, distinct from the simple linear relation expected from standard phase fluctuation scenario, $\lambda^{-2}$ vs $T_c$ must show a zero slope as $T_c$ approaches zero (since $\beta>0$).
This provides a natural explanation of the super-linear relation in the experimental observations shown in Fig.~\ref{fig4}(a) and (c).
In fact, with $\beta=1$ this formula describes very well the observed relation in the entire low-$T_c$ region in both underdoped and overdoped sides.
Not surprisingly, the resulting scaling $\lambda^{-2} \propto T_c^2$ agrees with that of the 3+1D $XY$-mode that describes quantum phase fluctuation of superfluidity upon suppressed kinetic energy~\cite{Franz,Hetel}.

Such an apparent ``symmetry'' near the quantum phase transition points between the underdoped and overdoped sides is puzzling within the current lore, given that one is supposedly governed by fluctuation in the phase and the other in the amplitude.
Especially, the observed physical properties of the normal-state above $T_c$ in these two regions are quite distinct~\cite{Damascelli,Dagotto,Basov}.
In our EBL picture, the contrast in non-superfluid properties is easily understood from the rising importance of the jamming effect discussed above.
Concerning the higher-energy features, Fig.~\ref{fig4}(d)-(f) show that in the overdoped region the renormalized $s$-wave (in blue) and $d$-wave bosons (in red) suffer a significant renormalization, quite distinct from the underdoped and optimally doped region.
Yet, the renormalized low-energy coherent $d$-wave bosonic band near the underdoped (d) and the overdoped transition points becomes heavier (flatter) in a similar fashion (through very different microscopic mechanism though.)
In other words, the low-energy phenomenon of superfluidity experiences a similar loss of phase stiffness due to the effective mass divergence despite the distinctly different underlying higher-energy physics.

From this perspective, the non-superconducting ``normal state'' above 25\% doping should have interesting properties.
While energetically most favorable locally, the pure $d$-wave boson would suffer from severe jamming and loss of kinetic energy.
It would thus tend to morph into a $p$-wave boson whose cigar shape allows them to remain mobile at a much larger doping (See Appendix~\ref{Beyond_overdope}).
Therefore, the EBL is expected to behave as an unconventional Bose metal consisting of fluctuating $d$- and $p$-wave bosons.
Alternatively, since at very high doping, the electronic correlation will eventually become weaker, the emergent bosons might lose their binding strength and start to decompose into weakly bound fermions.
Concerning this scenario, we note two very relevant experimental observations: 1) robust paramagnon dispersion at 40\% doping~\cite{Dean} indicating that the local magnetic correlation remains strong and 2) linear resistivity at 30\% doping~\cite{Bozovic1} indicating that the normal states is still not a fermionic quasi-particle based ``normal metal''.
Both observations suggest that decomposition of the emergent boson is not immediate after the disappearance of superconductivity.

It is also interesting to realize that in every energy scale our EBL experiences important influence of the ``Mottness'' of the underlying doped holes through the \textit{suppression of their double occupation} at each Cu site due to large local interaction.
First, with the help of kinetic energy, it lays the foundation of near neighboring antiferromagnetic~\cite{Anderson1950} and bi-polaronic~\cite{Alexandrov} correlations that provide a strong tendency to bind doped holes into pairs~\cite{Hirsch,Anderson1705}.
Then, it forces the bound pairs to occupy two atomic sites and thus indirectly establishes $d$-wave form~\cite{WeiKu2}.
Finally, it produces the extended hard-core constraint of the EBL that ultimately induces jamming of low-energy $d$-wave bosons at high enough density.
This jamming in turn suppresses the superfluid phase coherence especially with higher density.

Since the dominant physics of jamming inherits from the Mottness at very high energy scale, details of the intermediate- and low-energy scale is therefore not as essential, as long as the generic assumptions of our EBL model remains valid.
Interestingly, the recently discovered Nd$_{1-x}$Sr$_x$NiO$_2$ also demonstrates a superconducting critical doping at around 25\%~\cite{Danfeng1,Danfeng2}, similar to our analysis above.
This strongly suggests that the nickelate superconductors can also be described by an EBL, with strong jamming-induced phase fluctuation in the overdoped regime.
Future experimental verification of this prediction would be highly valuable.

In short, we show that emergent Bose liquid built from carriers under strong intra-atomic repulsion (Mottness) provides a generic mechanism that reduces superfluid stiffness at higher density due to jamming of its kinetic process at around 25-30\% carrier density.
This mechanism naturally explains the recent observation of strong phase fluctuation in the overdoped cuprates~\cite{Bozovic1,Bozovic2} and likely to be found also in the nickelates.
Assuming that this mechanism is indeed responsible for the puzzling observation in the cuprates, it further suggests that in the entire doping range of the superconducting dome, the unconventional superconductivity can be described by bosonic superfluidity for example of an EBL, without resorting to a crossover into BCS-like amplitude-fluctuating descriptions.

\section{Key characteristics of the boson in the EBL model}
\label{EBL_uniqueness}
The EBL model assumes a high-energy binding between the nearest neighboring carriers such that the unbinding processes can be integrated out from the low-energy Hilbert space.
Therefore, the bosons in EBL model are in a tightly bound two-body state of fermionic carriers of the nearest neighboring position (instead of momentum) with energy spanning far away from the chemical potential (since the binding is assumed strong). Most importantly, with strong short-range magnetic correlation in the background, the two-body state is not necessarily a spin singlet. 
Due to their emergent nature, they contain the following key characteristics that distinguish them from other boson-like carriers proposed in the literature, such as bosonized Cooper pairs~\cite{Dzhumanov2000,Dzhumanov2016,Dzhumanov2019}, resonating valence bond (RVB)~\cite{Anderson1973,Fazekas1974,Anderson1987}:

1. They are in a tightly bound two-body state of fermionic carriers (not necessarily a spin singlet, given a strong short-range magnetic correlation.)

2. They consist of fermionic state with energy spanning far away from the chemical potential, since the binding is assumed strong.

3. They are small in size and have a well-defined location, with underlying fermions residing in the nearest neighboring position (instead of a fixed total momentum as in a Cooper pair).

4. They have a rod-like shape with two ends at nearest neighboring atoms.

5. Correspondingly, they reside in the bonds between atomic sites.

6. The bosonic lattice is therefore the bond lattice, different from the underlying fermionic lattice.

7. For systems of 2D or higher dimension, the multiple directions of bonds imply multi-orbital nature of EBL.

8. The bosons in EBL carry a limited pivoting motion, moving one end at a time. This paves ways to frustration in the kinetic energy and can therefore heavily renormalize the kinetic energy of the first bosonic band (to $\sim30$~meV for the cuprates.)

9. They inherit an extended hard-core constraint from the large on-site repulsion between the underlying fermionic carriers.

\section{Applicability of EBL to the cuprates}
\label{EBL_check}

Discovered almost thirty years ago\cite{Muller}, the exotic phenomenon of high-temperature superconductivity (HT-SC) in the cuprates still remains puzzling to researchers.
Conventional superconductivity, a state of the matter that shows no resistance in conducting current, is well described by the standard ``BCS'' theory\cite{BCS} via weakly bound ``Cooper pairs" that fluctuate in amplitude.
On the other hand, the HT-SC in cuprates shows qualitatively different behavior.
For example, in the weakly hole doped (``underdoped'') region, the isotope effect increases dramatically and yet the corresponding superconducting transition temperature $T_c$ decreases~\cite{Kresin} instead.
The observed superconducting gap $\Delta_0$ in the cuprates is typically significantly larger than the canonical value of twice the transition temperature $\sim2k_BT_c$ in BCS theory~\cite{J.K.Ren, N.Miyakawa}.
During the phase transition, the measured specific heat shows two clear kinks 10K apart~\cite{Toshiaki}, qualitatively different from a standard second-order phase transition from the BCS theory.
Furthermore, in the underdoped region, the low-temperature specific heat shows no $T^2$ contribution expected from the observed $d$-wave quasiparticles, but only a dominant $T^3$ instead~\cite{Toshiaki}.
In addition, the observed upper critical field $H_{c2}$ does not saturate at low temperature and sometimes even exceeds the Pauli limit~\cite{Ando}.
These qualitative distinct features indicate clearly that the HT-SC in the cuprates is of a different nature.

The assumptions of the EBL model are actually quite applicable given the strong high-energy spin and lattice correlations observed in the cuprates~\cite{Anderson1950,Alexandrov,Sawatzky1}.
Particularly, the EBL model is in good consistency with the recent observations of bosonic behaviors in the cuprates~\cite{Bollinger2011,Zhou2019,Wang2021}, and is obviously the most intuitive one to account for the observed dominant role of phase fluctuation in the superconducting phase transition of the cuprates~\cite{Emery1,Bozovic1}.

Below we revisit several recent studies that demonstrated persistent agreement of optical~\cite{Zeng2021}, transport~\cite{Hegg2021,Zeng2021}, quasi-particle~\cite{WeiKu1,Jiang,Yue2021}, and superconducting~\cite{WeiKu2} properties of EBL with corresponding experimental observations.
The fact that these successful applications of EBL are all via the same Hamiltonian with the same set of parameters strongly supports the scenario in which charge-related physics in the cuprates can be understood intuitively by a simple EBL.

\subsection{Doping dependent Superconducting gap}

\begin{figure}
	 \setlength{\belowcaptionskip}{-0.1cm}
	\begin{center}
		\includegraphics[width=6cm]{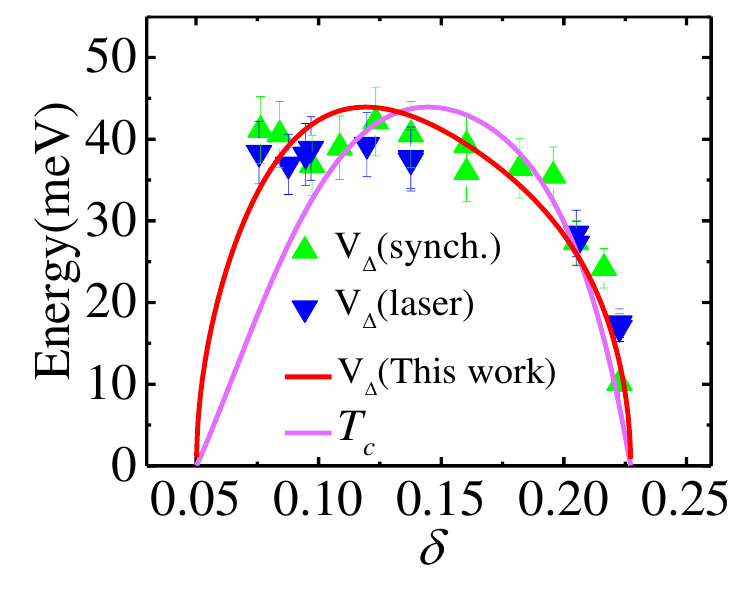}
	\end{center}
	\caption{Observed doping dependent superconducting gap size (triangles) deviates from superconducting temperature $T_c$ (purple line) but well captured by $\Delta\propto(1-2\delta)\sqrt{T_c}$ (red line).
    }
	\label{fig5}
\end{figure}

Previously this EBL model was shown to offer a simple explanation for the disappearance of superfluidity in cuprates at $\delta \sim 5\%$, as the effective mass of the emergent boson diverges~\cite{WeiKu2}.
(The same study also suggested a scenario for the lack of superfluidity below 5\% doping.)
In a related study, a second kind of ``superconducting gap'' was found in the quasi-particle spectrum, resulting from coherent \emph{kinetic} scattering against the Bose-Einstein condensation (BEC) of the EBL~\cite{WeiKu1}:
\begin{eqnarray}
\label{QPGap}
\Delta_{\mathbf{k}}(T)\sim f_d(k)\cdot (\epsilon_k-\epsilon_0)\sqrt{n_0(T)}
\end{eqnarray}
where $f_d(k)=\frac{1}{2}[\cos(k_x)-\cos(k_y)]$ gives the momentum dependent $d$-wave form factor of the order parameter and $n_0(T)$ the temperature-dependent condensation density.
$\epsilon_k-\epsilon_0$ denotes the quasi-particle energy measured from the low-energy band center, and signifies the kinetic origin of the gap (as opposed to the pairing potential).
Beyond the ``Fermi arc'', where $\epsilon_k$ moves from the chemical potential toward the band center, it provided additional momentum dependence~\cite{WeiKu1} that reproduced nicely the ARPES measurements~\cite{Kaminski} at various doping levels [cf. Fig.~\ref{fig1}(c).]

We first observe that this model actual ``predicted'' naturally the above-mentioned nearly doping independent superconducting gap in ARPES measurements~\cite{Vishik,Shen}.
Since the quasi-particles on the Fermi arc are at the chemical potential $\mu$, $\epsilon_k - \epsilon_0$ is given by $\mu - \epsilon_0 \propto 1 - 2\delta$ upon hole doping into the quasi-2D singlet band of the cuprates.
Away from the ends of the dome, in the absence of strong quantum fluctuation, the condensation density $n_0(T=0)$ is roughly proportional to the superfluid density, $n_s(T=0) \propto T_c$.
Together, this gives a weakly doping dependent near-node gap scale $V_\Delta \propto (1-2\delta) \sqrt{T_c}$, very similar to the experiments shown in Fig.~\ref{fig5}, but in great contrast to the stronger dome shape of $T_c$.
Particularly, notice that the $1 - 2\delta$ factor shifts the maximum of $V_\Delta$ to a significantly lower doping $\delta\sim0.12$, compared to $T_c$.
Similar to the above-mentioned deviation from $d$-wave form of the gap, this again reflects the kinetic nature of the anomalous scattering gap absent in typical weak-coupling pairing scenario.

\subsection{Transport, quasiparticle and superconducting properties}

Here, we recite our recent studies of various non-trivial physical properties of EBL in its non-superfluid and superfluid phase, which all demonstrate excellent agreement with experimental observations of the cuprates, within a \textit{single} set of parameters and the exact \textit{same} Hamiltonian of EBL.

First, the transport properties of EBL in the non-superfluid phase are generically non-Fermi liquid like, not surprisingly.
Its optical conductivity~\cite{Zeng2021} contains a featureless continuum extending a broad ($>300$ meV) energy range, with a mid-infrared peak around 100 meV, nicely reproducing the puzzling experimental spectra.
The corresponding DC resistivity~\cite{Zeng2021} demonstrates no saturation at high temperature (the ``bad metal'' behavior), a linear temperature dependence at a large temperature range all the way to very low temperature (the ``strange metal'' behavior), and even sometimes an insulator-like upturn at lower temperature (the ``weak insulator'' behavior).
These exotic non-Fermi liquid transport behaviors all exist at different temperature range of the same system, just like what is observed in the cuprate samples.
When $\tau^{\prime\prime} > \tau^\prime$ (corresponding to extremely underdoped $\delta < 5\%$ cuprates), EBL has been numerically~\cite{WeiKu2} and analytically~\cite{Hegg2021} proven to be a homogeneous Bose metal that cannot superflow and acquires finite resistivity with temperature and/or disorder without the protection of an energy gap, resembling the low-temperature pseudogap phase of the hole doped cuprates.
Under moderate magnetic field, not surprisingly EBL displays no quantum oscillation (since it has no Fermi surface).
The corresponding Hall coefficient is highly temperature dependent just like the cuprates observation, and does not give directly the carrier density~\cite{Lang2022}.

Second, upon scattering against EBL, the residual fermionic quasi-particle (QP) demonstrates strong non-Fermi liquid behaviors as well.
The EBL density of state contains a clear $\sim50$ meV structure and a large continuum extended to 300 meV~\cite{Jiang}, resembling perfectly the Eliashberg function $\alpha^2F$ extracted from ARPES.
The corresponding QP self-energy demonstrates two well-defined features at $\sim25$ meV (that render low-energy QP heavier) and $\sim50-70$ meV (that generates a ``kink'' in the QP dispersion), in excellent agreement with ARPES observation.
Due to the non-vanishing density of the EBL at low temperature (contrary to phonon and magnon), the normal-state QP demonstrates non-Fermi liquid scattering rate~\cite{Jiang,Yue2021} at the chemical potential, just like the ARPES observations.
Furthermore, since it is the same set of bosons the QP scatters against in the non-superfluid and the superfluid phase, the incoherent normal-state scattering rate is trivially proportional to the low-temperature coherent superconducting QP gap~\cite{Jiang}, explaining the ARPES observed puzzling connection between the incoherent and the coherent features.
Specifically in the Bose metal phase that corresponding to the pseudogap phase, scattering against EBL generates a Fermi arc with an obvious pseudogap~\cite{Yue2021} in the anti-nodal region that fills up (instead of closes) at higher temperature, same as the  observations by ARPES and scanning tunneling spectroscopy (STS).
For fixed momentum, the resulting pseudogap displays a strong asymmetric gap edge in energy, while for a fixed site it has a perfect symmetric gap edge~\cite{Yue2021}, reconciling the puzzling qualitative inconsistency between ARPES and STS.

Third, in the superfluid phase, coherent scattering against EBL generates a \textit{second} kind of QP superconducting gap~\cite{WeiKu1} whose energy scale corresponds to the effective kinetic energy, instead of the pairing energy.
This generates additional momentum dependence beyond that of the $d$-wave superconducting order parameter, similar to the weight transfer observation by ARPES.
Furthermore, this \textit{analytically and quantitatively} reproduces the correspondence between the QP dispersion and the superconducting gap observed by ARPES.

Finally, at low temperature, EBL was shown to host a $d$-wave superfluid~\cite{WeiKu2}, scattering against which the QP display a nodal direction between the second neighboring Cu-Cu directions~\cite{WeiKu1}, consistent with the observation.
The EBL superfluidity suffers from strong phase fluctuation associated with the mass divergence~\cite{WeiKu2} due to a local level crossing at $\tau^\prime \rightarrow\tau^{\prime\prime}$, corresponding to the observed quantum fluctuation near 5\% hole doping.
And, in this article, we show that a similar mass divergence has to take place at around 25-30\% carrier density, but this time due to the robust jamming effect induced by the Mottness of the electronic structure.
This effect leads to an unusual physical trend of weaker superconducting phase stiffness at higher density, as recently observed experimentally.

We stress that these persistent successful applications of the EBL model are all via the exact same Hamiltonian with the same set of parameters as the ones used in this article.
Combined with recent observations of bosonic behaviors in the cuprates~\cite{Bollinger2011,Zhou2019,Wang2021}, these results lay strong support for a consistent EBL-based scenario for the cuprates.
Further experimental confirmations, for example direct observation of the bosonic quasi-particles and their dispersion [c.f. Fig.~\ref{fig2}(a)] via one-photon-in-two-electron-out photo-emission process, or observation of the EBL-predicted second superconducting dome~\cite{Hegg2021} below 5\% doping under a strong (110) uniaxial pressure, would offer stronger verification (or falsification) of the suitability of the EBL in describing the low-energy physics of charges in the cuprates.

\section{Summary}

In summary, within the scenario emergent Bose liquid with extended hard-core constraint, we propose a generic mechanism to explain the puzzling reduction of superfluid phase stiffness and transition temperature at high carrier density.
Essentially, the strong intra-atomic repulsion (the Mottness) leads to a large extended hard-core constraint for the emergent bosons that can be easily jammed in their motion at high density and thereby host weakened superfluid stiffness.
Our results are well consistent with the experimental observations in some superconductors, such as the cuprates and the nickelates.
Furthermore, together the persistent consistency of the EBL with many key spectral and transport observations of the cuprates, this new finding adds credibility to the notion that EBL model can provide a good description for the charge channel of the low-energy electronic structure of the cuprates.

Note added: After initial submission of our manuscript, the shot noise experiment proposed in our original manuscript was performed~\cite{Zhou2019}, whose data indicated indeed a 2e-charge quanta, subject to further independent experimental confirmation.  In the final version of this manuscript, we therefore remove this suggestion and instead incorporate this new development in the reference.

\section{Acknowledgments}
This work is supported by National Natural Science Foundation of China (NSFC) under Grant Nos. 12274287 and 12042507, and Innovation Program for Quantum Science and Technology No. 2021ZD0301900. FY acknowledges support from NSFC No. 12074031 and 11674025.

\appendix
\section{Consideration of low-energy Fermions}
\label{App1}

\begin{figure}
	\begin{center}
		\includegraphics[width=\columnwidth]{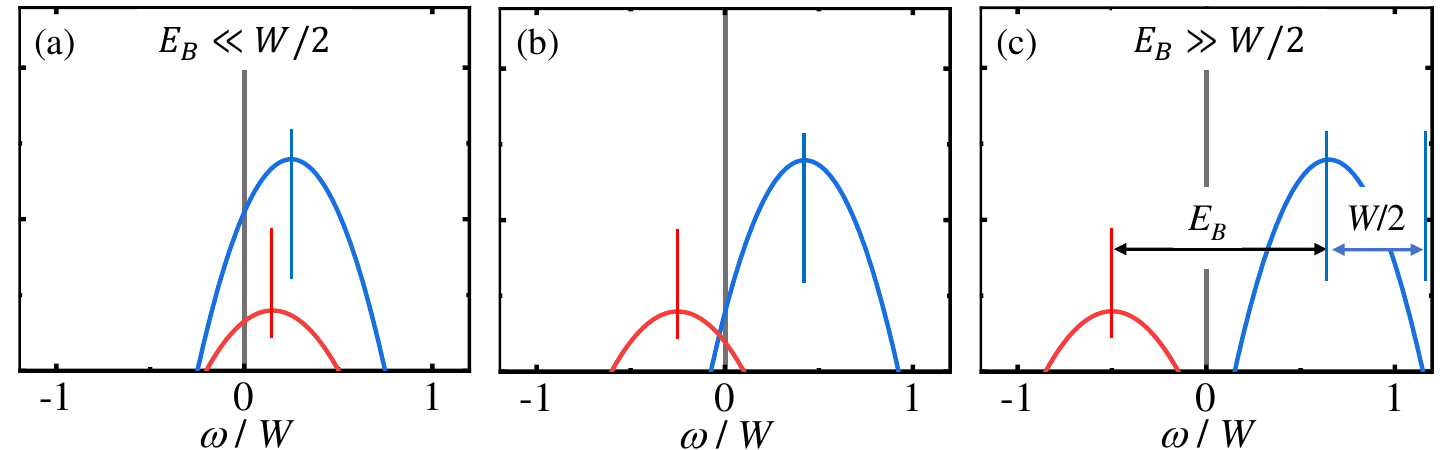}
	\caption{A schematic of the fermionic one-body spectral function (or density of states) under different binding strengths in unit of the bandwidth $W$ of unbound fermions. (a) In the weak-binding limit, the dominant occupied spectral weight (those below the chemical potential set at 0) come from unbound states given in blue. (b) With intermediate $E_B$ of the order of $W/2$, the fermions in the system are mostly in bound pairs in red, but still have small probability to be in the unbound state in blue that forms the Fermi surface. (c) In the strong-binding limit (the BEC limit) fermions are all in the bound state, leaving no unbound fermion at low energy below a clean gap scale of the order of $E_B - W$.
	}
	\label{spectrum}
	\end{center}
\end{figure}

It's a common question whether one is allowed to employ a bosonic picture like the EBL model to systems with experimentally observed low-energy fermionic quasi-particles, such as the cuprates~\cite{Damascelli,Fischer2007}.
As shown in Figure~\ref{spectrum}(c), if one would consider \textit{only} the BEC limit corresponding to an overwhelming binding energy $E_B$, the low-energy fermions indeed should be completely depleted with a clean gap of the scale of $E_B-W$ around the chemical potential in the fermionic one-body spectral function.
(Here $W$ denotes the bandwidth of unbound fermions.)
However, in general, with intermediate binding strength ($E_B$ slightly larger than $W/2$) illustrated in Figure~\ref{spectrum}(b)~\cite{Kuleeva2014} one often finds a small probability of the unbound fermions (in blue) being occupied and forming the Fermi surface.
Even though, fermions in the system are actually predominately part of the well-defined bound bosonic state, as reflected by the larger spectral weight in red below the chemical potential.
In this scenario, the rare gapless fermionic quasi-particles' contributions to low-energy physical properties are generally overwhelmed by those of the more likely bosonic carriers.
Thus, a bosonic description of the low-energy physics, such as the EBL model used in this study, would be the most convenient even in the presence of rare gapless unbound fermions.

Importantly, in contrast to the weak influence of the rare unbound fermions on the bound fermion, the latter is however expected to have significant impact on the former.
Particularly at low temperature, contrary to the depletion of typical bosonic excitations such as phonon and magnon, the number (or more precisely the probability) of these bound bosons is basically fixed.
Therefore, scattering against the bosonic carrier can lead to many unusual non-Fermi-liquid behaviors of the rare unbound fermions that cannot result from phonon or magnon, such as kink structure~\cite{Jiang}, Fermi arc and pseudogap~\cite{Yue2021} (c.f. Section~\ref{EBL_check} of the main text.)
In essence, the unbound fermions can sensitively reflect the influence of bosonic carriers and serve as good ``probes'' to the more essential bosonic carriers.

\section{Construction of the lowest-energy orbital and the corresponding one-orbital Hamiltonian}
\label{App2}
Since the focus of this study is on the superfluid stiffness, our main interest is the lowest-energy fluctuation around the superfluid.
We therefore aim to construct a low-energy quasi-particle whose local structure absorbs the influence of the high-energy orbitals.
Conceptually, this is similar to integrating out $\tau^\prime$ to construct a local Wannier orbital with a procedure similar to that of the Zhang-Rice singlet, where the nearest neighbor hopping processes between two oxygen are integrated~\cite{Zhang1988}.
As long as the energy of resulting local object is lower enough than other state, this approximation is controlled.

In practice, it is easily achieved numerically by simply constructing the $d$-wave Wannier function from the lower-energy band in Fig.~\ref{fig2}(a).
The resulting energy of $d$-wave boson is $\epsilon_d=\tau^{\prime\prime}-2\tau^\prime=-41.2$meV which is lower than the other states, such as $p$-wave boson with energy $\epsilon_p=-\tau^{\prime\prime}=-25.8$meV (with the parameter $\tau^\prime=33.5$meV, $\tau^{\prime\prime}=25.8$meV, see Table~\ref{tabA1}).
With the Wannier orbital, the lowest-energy effective one-band Hamiltonian Eq.~\ref{Honeband} can be obtained by representing Eq.~\ref{model1} within the subspace of $d$-wave boson, similar in spirit to the construction of the $t$-$J$ model from the Zhang-Rice singlet~\cite{Zhang1988}.

\section{Parameters of EBL Hamiltonian}
\label{parameters}

The key physics of interest in this study, namely enhanced superfluid phase fluctuation at higher density, is dominated by the extended hard-core constraint inheriting from the high-energy physics of Mottness.
Therefore, it is completely independent of the parameters of Eq.~\ref{model1}.
Nonetheless, this section provides the parameters for interested readers.

Table~\ref{tabA1} lists the parameters of Eq.~\ref{model1} used in this study.
These parameters were initially extracted~\cite{WeiKu1} from ARPES experiments~\cite{Yoshida2006,Yoshida2009} on La$_{2-\delta}$Sr$_\delta$CuO$_4$, and kept \textit{fixed} in all the studies~\cite{WeiKu1,WeiKu2,Jiang,Hegg2021,Yue2021,Zeng2021,Lang2022} using the same EBL model.
Other families of cuprates might have slightly different parameters, but we don't expect any significant deviation.
\begin{table}
    \caption{ARPES-extracted~\cite{WeiKu1,Yoshida2006,Yoshida2009} first and second nearest neighboring hopping parameters $\tau^\prime$ and $\tau^{\prime\prime}$ of bosonic carriers in Eq.~\ref{model1} at different doping level $\delta$ of La$_{2-\delta}$Sr$_\delta$CuO$_4$.
    Note that these parameters are kept fixed in all existing studies~\cite{WeiKu1,WeiKu2,Jiang,Hegg2021,Yue2021,Zeng2021,Lang2022} of this EBL model.}
   \begin{ruledtabular}
       \begin{tabular}{lllll}
         $\delta$ & 5.2\% & 7\% & 15\% & 22\%          \\\hline
$\tau^\prime$ (meV) & 29.8                    & 30.6        & 33.5      & 35.1      \\
$\tau^{\prime\prime}$ (meV) & 29.8                     & 29.0      & 25.8      & 23.9\\ 
\end{tabular}
    \end{ruledtabular}
    \label{tabA1}
	\vspace{-0.4cm}
\end{table}

Table~\ref{tabA2} gives examples of $\tilde{\tau}_{ii^\prime}$ of the $d$-wave boson at $\delta=15\%$, corresponding to the band in Fig.~\ref{fig2}(b).
They are derived from integrating out the $s$-wave boson corresponding to the higher-energy band in Fig.~\ref{fig2}(a).
\begin{table}
    \caption{Examples of kinetic parameters $\tilde{\tau}_{ii^\prime}$ of $d$-wave boson in Eq.~\ref{Honeband} for $\delta=$15\%.}
    \begin{ruledtabular}
       \begin{tabular}{llllll}
         $\mathbf{i}-\mathbf{i^\prime}$ & (1,0) &  (1,1)& (2,0)&(2,1) &(2,2)        \\\hline
$\tilde{\tau}_{ii^\prime}$ (meV) & -5.19                    & -6.02        & 3.61 &1.20 &-0.24  \\
\end{tabular}
    \end{ruledtabular}
    \label{tabA2}
	\vspace{-0.4cm}
\end{table}

\section{Calculation of Gutzwiller $g$-factor}
\label{gfac}
\subsection{Statistical counting of Gutzwiller $g$-factor}
We perform the statistical counting of the Gutzwiller $g$-factor $g_{ii\prime}$ numerically.
In a $M=10\times 10$ square lattice for $d$-wave bosons, we randomly place $N\equiv\frac{\delta}{2}M$ bosons and keep only the configurations that satisfy the 13-site extended hard core constraint described in the main text.
Within these valid configurations, we then count the probability $P$ of ``legal'' hoppings for each first, second, and third neighboring hopping $d_{i}^\dagger d_{i\prime}$ that do not lead to a violation of the constraint.

This probability is closely related to the Gutzwiller $g$-factor~\cite{Gutzwiller}, except that it ignores details of the lowest energy states.
Specifically, the definition of the Gutzwiller $g$-factor involves the ratio of thermal average of the hopping process
$g_{ii\prime}=\langle d_i^\dagger d_{i\prime}\rangle_c / \langle d_i^\dagger d_{i^\prime}\rangle_0$
evaluated using low-energy states with and without the constraint (denoted by subscripts $c$ and $0$.)
In our case, $d^\dagger_i$ is the creation operator of local $d$-wave boson.
Therefore, this $g$-factors should in general be energy and temperature dependent, especially around the scale of the interaction that induces this constraint.
On the other hand, below this energy scale where the constraint is enforced, the $g$-factor should be quite energy- and temperature-independent, and thus not very sensitive to the detailed structure of the low-lying states.
This makes such statistical counting a rather good approximation of the actual $g$-factor.
Indeed, as shown in the main text, the resulting $g$-factor resembles the VMC calculation very well.
This energy independence (within the scale of our low-energy Hamiltonian) also allows us to apply on an approximate ground state in our VMC calculation.

\subsection{Variational Monte Carlo (VMC) calculation of Gutzwiller $g$-factor}
Due to the extremely large many-body Hilbert space of bosonic systems, we can only afford to evaluate the Gutzwiller $g$-factor using an approximate wave function.
Luckily, as argued above, the results should be quite insensitive to the choice of wave function. 

We calculate the ratio between the expectation values of the hopping term $\langle d_{i}^\dagger d_{i\prime}\rangle$ in the ``Gutzwiller-projected" Bose-Einstein condensation (BEC) state and that in the mean-field BEC state
as the Gutzwiller $g$-factor, $g_{ii\prime}$,
\begin{equation}
g_{ii\prime}=\frac{\langle BEC \left|P_G d_{i}^\dagger d_{i\prime}P_G\right|BEC\rangle}{\langle BEC\left| d_{i}^\dagger d_{i\prime}\right|BEC\rangle}.
\label{definition}
\end{equation}
Here, the $P_G$ represents the extended hard-core constraint imposed on the bosonic hole-pairs, and the BEC mean-field state $|BEC\rangle$ is expressed by the following normalized formula,
\begin{equation}
\label{BEC}
|BEC\rangle\equiv\frac{1}{\sqrt{N!}}(d^{\dagger}_{\mathbf{k}=0})^{N}|0\rangle=\frac{1}{\sqrt{N!}}(\frac{1}{\sqrt{M}}\sum_{i=1}^{M}d^{\dagger}_i)^{N}|0\rangle,
\end{equation}
where $M$ and $N$ represent the total site number and boson numbers respectively, with $N=M\delta/2$.

The denominator of Eq.(\ref{definition}) can be easily obtained as,
\begin{equation}
\label{denorminator}
\begin{split}
    \langle BEC\left| d_{i}^\dagger d_{i\prime}\right|BEC\rangle&=\frac{1}{M}\sum_{\mathbf{k}}e^{-i\mathbf{k}\cdot(\mathbf{R}_i-\mathbf{R}{_i\prime)}}\langle BEC\left| d_{\mathbf{k}}^\dagger d_{\mathbf{k}}\right|BEC\rangle\\
    &=\frac{1}{M}\langle BEC\left| d_{0}^\dagger d_{0}\right|BEC\rangle=\frac{N}{M}=\delta/2.
\end{split}
\end{equation}
The numerator of Eq.(\ref{definition}) can be evaluated as,
\begin{equation}
\label{numerator}
\begin{split}
    &\langle BEC\left|P_Gd_{i}^\dagger d_{i\prime}P_G\right|BEC\rangle\\
    &=\sum_{\alpha}\left|\left\langle\alpha\left|P_G\right| BEC\right\rangle\right|^2\sum_{
\beta}\left\langle\alpha\left|d_{i}^{\dagger}d_{i\prime}\right|\beta\right\rangle\cdot\frac{\left\langle\beta\left|P_G\right| BEC\right\rangle}{\left\langle\alpha\left|P_G\right| BEC\right\rangle}\\
&\equiv\sum_{\alpha}P_{\alpha}B_{\alpha},
\end{split}
\end{equation}
where $P_{\alpha}\equiv\left|\left\langle\alpha\left|P_G\right| BEC\right\rangle\right|^2$ represents the weight of each configuration $\left|\alpha\right\rangle$ in the Gutzwiller-projected wave function $P_G|BEC\rangle$, and $B_{\alpha}=\sum_{
\beta}\left\langle\alpha\left|d_{i}^{\dagger}d_{i\prime}\right|\beta\right\rangle\cdot\frac{\left\langle\beta\left|P_G\right| BEC\right\rangle}{\left\langle\alpha\left|P_G\right| BEC\right\rangle}$ represents the measurement for the configuration $\left|\alpha\right\rangle$. 
The weight $P_{\alpha}$ and the measurement $B_{\alpha}$ for any configuration $|\alpha\rangle$ are easily obtained. From Eq.(\ref{BEC}), one easily finds that the weights $P_{\alpha}$ for all the configurations $|\alpha\rangle$ which are permitted by the extended hard-core constraint are equal, while those for the constraint-prohibited configurations are zero. 
The value $B_{\alpha}$ of any constraint-permitted configuration $|\alpha\rangle$ is chosen as follows: defining $|\beta\rangle=d_{i\prime}^{\dagger}d_{i}|\alpha\rangle$, if configuration $|\beta\rangle$ is permitted by the extended hard-core constraint then $B_{\alpha}=1$; otherwise $B_{\alpha}=0$. These formulae provide an appropriate start-point for the following Monte-Carlo calculations.

In the Monte-Carlo calculation, we start from an arbitrary configuration $|\alpha_1\rangle$. 
We then randomly select a particle in configuration that let the particle hop to any other hard-core-constraint-permitted position on the lattice to obtain a second configuration $|\alpha_2\rangle$. This completes an update. Continuing the updates, we obtain a configuration series $|\alpha_1\rangle,|\alpha_2\rangle,|\alpha_3\rangle,\cdots\cdots$. After about $10^{4}$ Monte-Carlo steps thermalization can be realized. 
In successive Monte-Carlo steps, one begins to extract the measurement $B_{\alpha}$ for the configuration $|\alpha\rangle$, and to average these $B_{\alpha}$ to obtain the numerator defined by Eq.(\ref{numerator}).
To avoid auto-correlation, we take one measurement after each N Monte-Carlo steps. About $10^{5}$ measurements are performed to attain convergence. 
The final result for the Gutzwiller $g$-factor is given by Eq.(\ref{definition}), Eq.(\ref{denorminator}) and Eq.(\ref{numerator}).


\section{Beyond overdoped system: possible Bose metal consisting of mixture of $p$- and $d$-wave EBL}
\label{Beyond_overdope}

\begin{figure*}
    \centering
    \includegraphics[width=15cm]{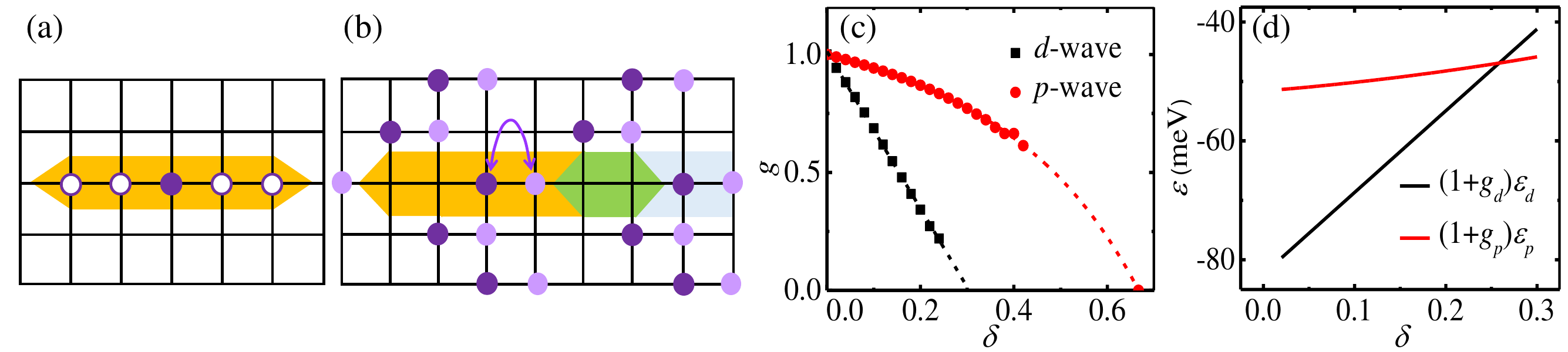}
    \caption{(a) Extended hard-core constraint of site-centered $p$-wave boson. (b) Simple estimation of the highest density of $p$-wave boson with minimal non-zero kinetic energy. (c) Gutzwiller $g$-factor of pure $p$- and $d$-wave boson for $\tau^\prime$ calculated by statistical counting. (d) Average kinetic energy of $d$- and $p$-wave boson modified by jamming effect, estimated with a fixed $\tau^\prime$ and $\tau^{\prime\prime}$.  At $\delta\sim 25\%$, the energy of $p$-wave boson becomes lower than $d$-wave boson.}
    \label{figs1}
\end{figure*}

Since we are interested in the superfluid phase stiffness in this study, our main study focuses only on the jamming of the $d$-wave emergent bosons.
In this section we briefly discuss the possible phase beyond the overdoped superfluid regime, where the jamming of the $d$-wave emergent bosons costs them too much kinetic energy.

Let's first recall that at low density, the freely propagating $d$-wave bosons gain the most kinetic energy, $2\epsilon_d=2\tau^{\prime\prime}-4\tau^{\prime}$, compared to those of other local symmetries.
This energy can be approximately split into two contributions: locally forming a $d$-wave form [a four-site Wannier function in Fig.~\ref{fig2}(b)] and its propagation in the system.
The former contribution can be easily estimated by solving a local four-site problem and turns out to be $\epsilon_d$, exactly half of the total kinetic energy.
Similarly, a $p$-wave emergent boson would have a total kinetic energy $2\epsilon_p=-2\tau^{\prime\prime}$, half of which is also gained by forming local $p$-wave symmetry.
Therefore, when $\tau^\prime > \tau^{\prime\prime}$, $2\epsilon_d < 2\epsilon_p$ and $d$-wave form is dominant.

As discussed in the main text, as the density increases upon higher doping, due to the large hard core the $d$-wave boson starts to suffer from a significant jamming effect.
Represented by the Gutzwiller factor $g_d$, a crude estimation of the jamming-induced renormalized kinetic energy can be made: $(1+g_d)\epsilon_d$.
Similarly, for the $p$-wave boson, the energy reduces to $(1+g_p)\epsilon_p$.

Interestingly, as shown in Figure.~\ref{figs1}(a) due to the narrower shape of the $p$-wave bosons, they naturally suffer a lot less from the jamming effect than the $d$-wave bosons.
Consequently, the $p$-wave boson can survive a much larger doping, approximately $\delta=50\%$ based on the estimation in Figure.~\ref{figs1}(b).
Indeed, Figure.~\ref{figs1}(c) shows a much slower decay in $g_p$ against $\delta$ from a simple statistical counting.
Therefore, one can expect that at high-enough doping, the $d$-wave symmetry should no longer be dominant, since the bosons would be forced to adapt to $p$-wave whenever they need to pass each other in short distance.
Figure.~\ref{figs1}(d) shows that for a fixed $\tau^\prime$ and $\tau^{\prime\prime}$ (from optimal doping), $(1+g_p)\epsilon_p$ will eventually become lower than $(1+g_d)\epsilon_d$.
Therefore, assuming that the emergent Bose liquid remains intact beyond the superconducting dome, the system should become a Bose metal consisting of a mixture of $p$- and $d$-wave bosons.
Notice that such a Bose metal should experience a different structure of main phase fluctuation compared with the recently proven~\cite{Hegg2021} homogeneous Bose metal in the underdoped region, and is an interesting candidate for another homogeneous Bose metal.
Such a strong phase fluctuating EBL should present non-Fermi liquid behaviors as well, as found experimentally in Ref.~\cite{Bozovic1,Dean}.
The emergence of $p$-wave boson is also consistent with the observation of a similar direction of the charge order found in beyond underdoped and beyond overdoped systems~\cite{wang}.

\bibliography{MainTex}

\begin{thebibliography}{64}
\bibitem [{dou()}]{doubleoccupacy}%
  \BibitemOpen
  \href@noop {} {}\bibinfo {note} {The non-double occupancy for electrons
  dictates at most one electron for each site in the parent compound. It is
  therefore impossible to inject 2 doped holes into one site, thus the double
  occupancy constraint of the doped holes.}\BibitemShut {Stop}%
\bibitem [{BEC()}]{BEC}%
  \BibitemOpen
  \href@noop {} {}\bibinfo {note} {Although the fermionic carriers in the
  cuprates have quasi-2D kinetic hopping, the low-temperature superfluid is
  actually 3D, with superfluid density in all directions vanishing at the same
  transition temperature. Interestingly, our emergent bosonic carriers also
  have 3D kinetic hopping, due to their large mass enhancement in the $ab$
  plane in association with the local interference between the hopping
  processes~\cite{WeiKu2}.}\BibitemShut {Stop}%
\end{thebibliography}
\end{document}